\documentclass[prl,aps,twocolumn,showpacs,preprintnumbers,amsmath,amssymb]{revtex4}
\usepackage{graphicx}
\usepackage{amssymb}
\usepackage{amsmath}
\usepackage{amsfonts}
\usepackage{bm}
\begin{document}
\title{Experimental verification of statistical correlation for bosons: Another kind of Hong-Ou-Mandel interference}
\author{Wei-Tao Liu$^1$}
\email{mugualaw@hotmail.com}
\author{Wei Wu$^1$}
\author{Ping-Xing Chen$^{1,2}$}
\email{pxchen@nudt.edu.cn}
\author{Cheng-Zu Li$^1$}
\author{Jian-Min Yuan$^1$}
\affiliation{1 Department of Physics,
National University of Defense Technology, Changsha, 410073, China\\
2 State Key Laboratory of Precision Spectroscopy, East China Normal
University, Shanghai 200062, China}
\date{\today}
\begin{abstract}
According to the identity principle in quantum theory, states of a
system consisted of identical particles should maintain unchanged
under interchanging between two of the particles. The whole
wavefunction should be symmetrized or antisymmetrized. This leads to
statistical correlations between particles, which exhibit observable
effects. We design an experiment to directly observe such effects
for bosons. The experiment is performed with two photons. The effect
of statistical correlations is clearly observed when the wavepackets
of two photons are completely overlapped, and this effect varies
with the degree of overlapping. The results of our experiment
substantiate the statistical correlation in a simple way. Experiment reported here can also
be regarded as another kind of two-photon Hong-Ou-Mandel interference, occurs in the polarization degree of freedom of photon.
\end{abstract}
\maketitle
One of the basic
principles consolidating the foundation of quantum mechanics is
identity principle, which indicates that there will be no observable
change in a system consisted of identical particles when two of the
particles are interchanged with each other\cite{dirac}. To meet this
request, the whole wavefunction of the system should be symmetrical
for bosons, while antisymmetrical for fermions. This leads to a
quite different effect from classical physics. In a system
containing multiple identical particles, there should be
correlations between particles since only symmetrical or
antisymmetrical state are permitted. It is known as statistical
correlation. Many kinds of physical phenomena illuminate the
existence of statistical correlation, such as
superconductivity\cite{superexp,superth} and Bose-Einstein
condensate\cite{bec,becexp}. That kind of correlation is closely
related with indistinguishability of quantum states, which also
excited many interests in fundamental problems in quantum mechanics
and applications in quantum information processing\cite{qip,qip2}.

Statistical correlations result in observable effects. We demonstrate an experiment with photons to directly observe the effect of
statistical correlation in boson system, by engineering the overlapping between the
wavepackets of two photons and performing proper measurements. The
results verify the correlations between two photons when their
wavepackets are completely overlapped. The effect of correlation
also varies with the degree of overlapping between two photons.
By comparison, we find out the connection between our experiment and
that of Hong-Ou-Mandel (HOM) interference\cite{hom}: our experiment can be regarded as
another kind of HOM interference which occurs in the polarization
degree of freedom of photon, while HOM interference can also be explained in the way similar to our experiment.

The main idea of the experiment is shown as following. Consider two
independent photons, one of which is horizontally polarized while
the other is vertically polarized. Obviously, if two photons are
spatially separated, they are distinguishable, and the two-photon
polarization state should be a product state
\begin{equation}\label{pstate}
|\psi_s\rangle=|H\rangle|V\rangle\  \text{or}\ |V\rangle|H\rangle,
\end{equation}
with $|H\rangle$($|V\rangle$) represents the
horizontally(vertically) polarized state. What is the two-photon
polarization state when their wavepackets are overlapped in
temporal-spatial space? Since they are indistinguishable when they
overlap, statistical correlation between photons should be
considered according to identity principle. For photons which are
known as bosons, the whole wavefunction should be symmetrical under
interchanging between two photons. Since the temporal-spatial part
of the whole wavefunction is symmetrical, the polarization part of
the wavefunction should be symmetrized to ensure the symmetry of the
whole wavefunction. Thus we obtain the polarization state
\begin{equation}\label{estate}
|\psi_o\rangle=\frac{1}{\sqrt{2}}(|H\rangle|V\rangle+|V\rangle|H\rangle).
\end{equation}
It turns out to be a polarization entangled state.

To verify the effect of statistical correlation, i.e., to verify
that two-photon polarization state does be the entangled state shown
in Eq.(\ref{estate}) instead of a product state when two photons are
indistinguishable, what we need to do is to find out the difference
between the product state and the entangled state. Toward this,
quantum state tomography\cite{qip} can not work well here since
those processes are designed for separated photons. Here, we perform
direct projection measurement on the two-photon state. Projection on
the state $|D\rangle|D\rangle$ are considered, where
$|D\rangle=(|H\rangle+|V\rangle)/\sqrt{2}$. When two photons are
separated and their polarization state is shown in
Eq.(\ref{pstate}), the probability of the state to be projected on
$|D\rangle|D\rangle$ is
\begin{equation}\label{product}
P_s=|\langle H|\langle V|D\rangle|D\rangle|^2=\frac{1}{4}
\end{equation}
While two photons are completely overlapped, the state shown in
Eq.(\ref{estate}) can be rewritten in the basis $\{|D\rangle,
|A\rangle=(|H\rangle-|V\rangle)/\sqrt{2}\}$ as
\begin{equation}\label{dabasis}
|\psi_o\rangle=\frac{1}{\sqrt{2}}(|D\rangle|D\rangle-|A\rangle|A\rangle)
\end{equation}
Thus the probability of projection on the state $|D\rangle|D\rangle$
will be
\begin{equation}\label{entanglement}
P_o=|\langle\psi_o|D\rangle|D\rangle|^2=\frac{1}{2}.
\end{equation}
Therefore, the effect of statistical correlations can be observed
directly by measuring the probability of the state projection on the
state $|D\rangle|D\rangle$.
\begin{figure}[t]
  \includegraphics[width=0.45\textwidth]{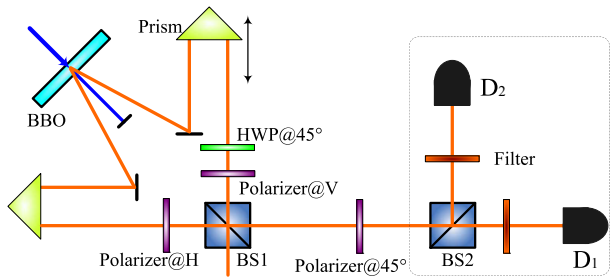}\\
  \caption{Experimental arrangements for directly observing the effect of statistical correlations with photons.
   Two photons are created via
  SPDC and firstly prepared in a product state of $|HV\rangle$. Then they are
  combined into a single beam with a beam splitter (BS1). A polarizer oriented at
  $+45^\circ$ is employed to perform the projection measurement on the state $|D\rangle|D\rangle$.
  The number of photon pairs is counted with another beam splitter (BS2)
  and two single photon detectors. The degree of overlapping
  between two photons is engineered by changing the length difference between two arms
  via scanning
  a prism located on a PC-controlled motor.}\label{exp}
\end{figure}

In the experiments, two photons are created via spontaneous
parametric down-conversion (SPDC). The experimental setup is shown
in Fig.\ref{exp}. A continuous wave laser (Coherent, MBR110 and
MBD200) operated at a wavelength of 397 nm with a power of 500 mW
serves as the pump source. A 0.59-mm-thick $\beta$-barium borate
(BBO) crystal cut for type-I degenerate noncollinear phase matching
is used as the down-converter. The down-converted photons are
detected with single photon detectors (PerkinElmer, SPCM-AQR-16). An
interference spectral filter centered at 794 nm with 10 nm bandwidth
precedes each detector, which result in a coherence time of
$\tau_c=210 fs$. The polarizer set at $45^\circ$ serves for the
projection measurement on the state $|D\rangle|D\rangle$.

Both photons are initially created in horizontal polarization state
and one of them is transformed into vertical polarization with a
half wave plate (HWP), the fast axis of which is set at an angle of
$+45^\circ$ with respect to the vertical direction. Two polarizers
respectively set to transmit horizontal and vertical polarization
states are employed to ensure the purity of the polarization state
of two photons.
\begin{figure}[b]
  \includegraphics[width=0.4\textwidth]{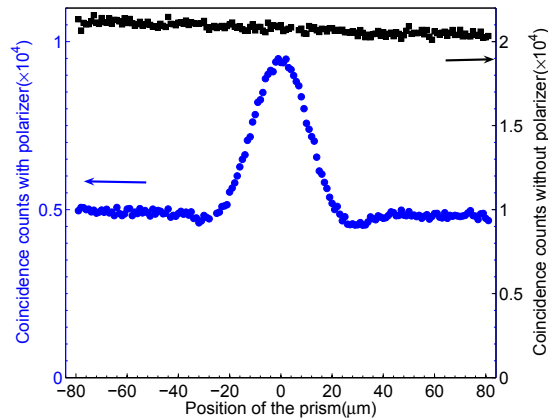}\\
  \caption{Coincident counts between two detectors with the prism scanning.
  Dark squares show the counts without projection measurement (Polarizer@$45^\circ$ removed).
  The round dots show the counts with projection measurement.}\label{result}
\end{figure}

Then two photons are combined with a polarization-independent 50:50
beam splitter (BS1) into a single beam, as shown in Fig.\ref{exp}.
There are four cases for two photons flying from two outputs of BS1.
Only those cases that two photons leave BS1 via the same one of the
outputs are considered, the probability of which is 1/4. The degree
of overlapping between the wavepackets of two photons is controlled
by changing the path length of one photon, which is performed with a
prism located on a computer-controlled motor. In the experiments,
path length difference between two photons are scanned from $-160\mu
m$ to $160\mu m$, thus the difference of arrival time between two
photons varies from -533 fs to 533 fs. When the path difference is
large enough, say $160\mu m$, two photons are temporally separated
since the difference of arrival time is larger than coherence time
of the photons. When the prism is located at the position of 0, two
paths are in the same length and two photons are completely
overlapped.

Firstly, we count the number of the photon pairs,
\emph{\textbf{without}} projection measurement. The photon pair
counting is performed with another beam splitter (BS2) and two
detectors, as shown in the dashed box in Fig.\ref{exp}. Only those
cases that both detectors click are recorded. Since detectors only
perform exist-or-not measurements and no polarization information of
the photons is revealed after photons going through BS2, the
post-selection of detection does not causing any interference. The
ideal efficiency of pair counting is 50\% regardless of the
efficiency of the detectors, whether there is correlation between
two photons or not. The results are shown as dark squares in
Fig.\ref{result}. The measured count of the photon pairs is
$\mathcal {N}=20777\pm308$, which shows no change whether two
photons are overlapped or not. The coincidence window is chosen to
be 10 ns for all the measurements.

Then the projection measurement is performed with a polarizer set at
$+45^\circ$ to transmit photons in the state $|D\rangle$, and photon
pairs transmit through the polarizer are also counted with the
photon-pair-counter shown in the dashed box in Fig.\ref{exp}. When
two photons are temporally separated, they are in product state
since they are distinguishable. For product state, the number of
photon pairs in state$|D\rangle|D\rangle$ should be $\mathcal{N}/4$
according to Eq.(\ref{product}), while it should be $\mathcal{N}/2$
when two photons overlap due to statistical correlation as shown in
Eq.(\ref{entanglement}). The results of the measurement varies with
the change of path length difference between two photons, shown as
round dots in Fig.\ref{result}. When two photons temporally
separated, the measured count is $4867\pm96$, almost one fourth of
$\mathcal{N}$, including the loss costed by the polarizer. Then a
maximal count of 9489 is obtained when two photons totally
overlapped, which is almost a half of $\mathcal{N}$.

Therefore, the data obtained in the experiments agree well with
Eq.(\ref{product}) and Eq.(\ref{entanglement}), which verifies the
effect of statistical correlation.
\begin{figure}[b]
  \includegraphics[width=0.45\textwidth]{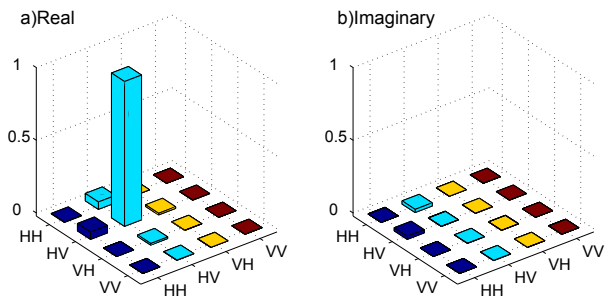}\\
  \caption{The reconstructed density matrix of the two-photon polarization state before they
  light on BS1, with the real part of
  the matrix on the left and the imaginary part on the right.}\label{HV}
\end{figure}

Statistical correlations are taking place not only when photons are
completely overlapped. From Fig.\ref{result}, the coincidence counts
continuously increase to the maximum when the difference of arrival
time between photons continuously decreasing to zero. That is, when
two photons are partially overlapped, correlations are also
observable, what's different is the degree of correlation. In our
experiments, only two photons are considered for convenience. For
the cases that more photons involved, it's possible to exhibit more
considerable effects.

To ensure that two photons traveling in the same beam,
Hong-Ou-Mandel interference\cite{hom} is observed in the
experiments. Two photons are transformed into the same polarization state with HWP and two polarizers shown in Fig.\ref{exp}, and Hong-Ou-Mandel
interference shows up with the prism scanning. The interference
visibility is observed to be higher than 97\%. Then the two photons are set back to be orthogonally polarized again for subsequent projection
measurement and Hong-Ou-Mandel interference
disappears\cite{idependent} under this situation.

In addition, the state of the two photons is an entangled state when
wavepackets of two photons are overlapped. This entanglement does
not exist before two photons being combined on BS1. To make sure
that, we perform quantum state tomography\cite{tomography} on the
polarization state of two photons before they light on BS1. The
results show that it indeed is a product state. The density matrix
reconstructed from the experimental data with a maximum likelihood
technique\cite{tomography} is shown in Fig.\ref{HV}. The
fidelity\cite{fidelity} between the measured density matrix and that
of the state $|H\rangle|V\rangle$ is 0.996.

That is, the state of two photons changes from a product state into
an entangled state when wavepackets of two photons are overlapped.
Generally speaking, entanglement can not be created without
interaction. However, there is no interaction between photons as is
well known. What makes two photons entangled? It is the equivalent
interaction of statistical correlations. When two photons are
overlapped with each other, the wavefunction has to be symmetrized,
which makes two photons correlated with each other. Therefore they
become entangled. Unfortunately, no applications on this kind of
entanglement and interaction has been discussed so far.

\begin{figure}[t]
  \includegraphics[width=0.4\textwidth]{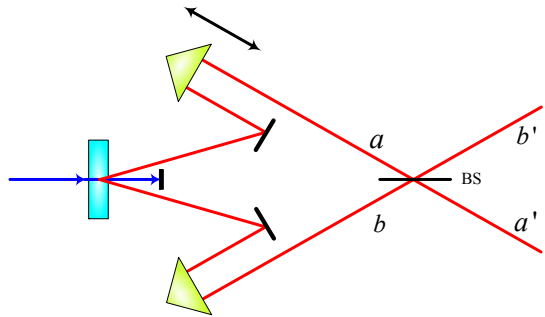}\\
  \caption{The diagram of Hong-Ou-Mandel interference experiment.}\label{hom}
\end{figure}

Now let's recall the experiment of Hong-Ou-Mandel interference and compare with our experiment.
Following the above reasoning used for our experiment, HOM interference can also be well explained.
In HOM experiment, two photons are prepared in different path states, as shown in Fig.\ref{hom},
one photon being in the state of $|a\rangle$ while the other in $|b\rangle$, with $\langle a|b\rangle=0$.
So the initial state of two-photon should be $|\psi_i\rangle=|a\rangle|b\rangle$. When two photons overlap on
the beam splitter, the two-photon state becomes to be $|\psi_e\rangle=(|a\rangle|b\rangle+|b\rangle|a\rangle)/\sqrt{2}$
according to statistical correlation between two photons. Represented in another basis $\{|a'\rangle,|b'\rangle\}$,
which contains the two output states of the beam splitter, with
$|a'\rangle=(|a\rangle+|b\rangle)/\sqrt{2}$ and $|b'\rangle=(|b\rangle-|a\rangle)/\sqrt{2}$. The
two-photon state can be rewritten as $|\psi_e\rangle=(|a'\rangle|a'\rangle-|b'\rangle|b'\rangle)/\sqrt{2}$.
Therefore when two photons leave the beam splitter, they will always travel out of the same output of BS. Coincidence
detection between two outputs of BS is equivalent to the projection on the state of $|a'\rangle|b'\rangle$ with a result
of zero, which agree with the results of Hong-Ou-Mandel interference. Similarly, if we project two-photon state on the state
$|a'\rangle|a'\rangle$ or $|b'\rangle|b'\rangle$, that is, detecting two photons at the same output of BS, the coincidence
count rate will increase when two photons overlap on BS\cite{sameout}.

On the other hand, our experiment can also be regarded as two-photon interference, similar to HOM interference.
When two photons leave BS1 in the same output, the
two-photon state can be written as $\hat{a}^\dag_H(t)\hat{a}^\dag_V(t+\tau)|00\rangle$, with $\hat{a}^\dag_H(t)$ being
the creation operator of a horizontal polarized photon at time $t$, and $\tau$ being the time delay between two photons. For photons in the states $|D\rangle$ or $|A\rangle$,
the corresponding creation operators are
\begin{align}
\hat{a}^\dag_D&=\frac{1}{\sqrt{2}}(\hat{a}^\dag_H+\hat{a}^\dag_V),\nonumber\\
\hat{a}^\dag_A&=\frac{1}{\sqrt{2}}(\hat{a}^\dag_H-\hat{a}^\dag_V).
\end{align}
Therefore, the two-photon state becomes
\begin{align}\label{interfer}
|\psi_o\rangle=&\frac{1}{2}\left[\hat{a}^\dag_D(t)+\hat{a}^\dag_A(t)\right]
\left[\hat{a}^\dag_D(t+\tau)-\hat{a}^\dag_A(t+\tau)\right]|00\rangle\nonumber\\
=&\frac{1}{2}\Bigl[\hat{a}^\dag_D(t)\hat{a}^\dag_D(t+\tau)-\hat{a}^\dag_A(t)\hat{a}^\dag_D(t+\tau)\nonumber\\
&+\hat{a}^\dag_D(t)\hat{a}^\dag_A(t+\tau)-\hat{a}^\dag_A(t)\hat{a}^\dag_D(t+\tau)\Bigr]|00\rangle.
\end{align}
When two photons completely overlap, $\tau=0$ and the last two terms in the above equation can not be distinguished which lead to destructive interference. The renormalized state is
\begin{equation}
|\psi_o\rangle=\frac{1}{\sqrt{2}}(\hat{a}^\dag_D\hat{a}^\dag_D-\hat{a}^\dag_A\hat{a}^\dag_A)|00\rangle
\end{equation}
which is the same as the state shown in Eq.(\ref{dabasis}).

In this sense, our experiment can be treated as another kind of HOM interference. The original HOM interference is occurs in the path degree of freedom while our experiment show two-photon interference in the polarization degree of freedom. Although these experiments can be explained in both ways, this two kinds of explanation do not contradict with each other. When the state is represented with creation operators acting on vacuum state, the effects of statistic correlations are involved. As shown in Eq.(\ref{interfer}),
the operators $\hat{a}^\dag_D$ and $\hat{a}^\dag_A$ commute with each other when $\tau=0$, which makes the last two terms indistinguishable and leads to completely destructive interference. For the cases that more photons involved, there will be more possible states involved and explaining with statistical correlations among photons will be more concise. In addition, when two photons in different polarization states from two different path overlap on a beam splitter, the whole wavefunction should be symmetrized according to identity principle, which can be used for Bell state analyzing for two-photon polarization states\cite{dense,pan}.

In conclusion, we designed an experiment to directly observe the
effect of statistical correlations for bosons. Two photons in
orthogonal polarization states were considered. By engineering the
degree of overlapping between two photons and performing projection
measurement on the photons, the effect of statistical correlations
is substantiated. The experiment can also be regarded as another kind of two-photon Hong-Ou-Mandel interference, which occurs in the polarization degree of freedom of photon.

The authors thank Prof. Zhe-Yu Jeff Ou and Prof. Guo-Xiang Huang for
useful discussions. This work is supported by National Natural
Science Foundation of China (Grant No. 10774192), Fund of
Innovation (No. B060204) from Graduate School of NUDT, and the opening
research foundation of State Key Laboratory of Precision
Spectroscopy.


\begin{thebibliography}{1}
\expandafter\ifx\csname bibnamefont\endcsname\relax
  \def\bibnamefont#1{#1}\fi
\expandafter\ifx\csname bibfnamefont\endcsname\relax
  \def\bibfnamefont#1{#1}\fi
\expandafter\ifx\csname citenamefont\endcsname\relax
  \def\citenamefont#1{#1}\fi
\providecommand{\bibinfo}[2]{#2} \providecommand{\eprint}[2]

\bibitem{dirac}P. A. M. Dirac, \emph{The Principles of Quantum Mechanics} (Oxford University Press,
1947).

\bibitem{superexp}H. K. Onnes, Comm. Phys. Lab. Univ. Leiden, Nos.
119, 120, 122(1911)

\bibitem{superth}J. Bardeen, L. N. Cooper, and J. R. Schrieffer,
Phys. Rev., \textbf{108}, 1175(1957)

\bibitem{bec}S. N. Bose, Zeitschrift f\"{u}r Physik, \textbf{26},
178(1924)

\bibitem{becexp}M. H. Anderson, J. R. Ensher, M. R. Matthews, C. E. Wieman, and E. A. Cornell,
Science, \textbf{269}, 198(1995)

\bibitem{qip}M. A. Nielsen and I. L. Chuang, \emph{Quantum Computation and
Quantum Information} (Cambridge University Press, Cambridge, UK,
2001).

\bibitem{qip2}N. Paunkovic, \emph{The Role of Indistinguishability of
Identical Particles in Quantum Information Processing}(Ph.D. thesis,
University of Oxford, 2004).

\bibitem{hom}C. K. Hong, Z. Y. Ou and L. Mandel, \prl, \textbf{59},
2044 (1987)

\bibitem{idependent}R. Kaltenbaek, B. Blauensteiner, M. \.{Z}ukowski, M. Aspelmeyer, and A.
Zeilinger, \prl, \textbf{96}, 240502(2006)

\bibitem{tomography}D. F. V. James, P. G. Kwiat, W. J. Munro, and A. G. White,
\pra, \textbf{64}, 052312 (2001).

\bibitem{fidelity}R. Jozsa, J. Mod. Opt. \textbf{41}, 2315 (1994)

\bibitem{sameout}J. G. Rarity and P. R. Tapster, J. Opt. Soc. Am. B,
\textbf{6}, 1221(1989)

\bibitem{dense}K. Mattle, H. Weinfurter, P. G. Kwiat and A. Zeilinger, Phys. Rev. Lett., \textbf{76}, 4656(1996)

\bibitem{pan}J. W. Pan, PhD thesis, Durchgef\"{u}hrt
am Institut f\"{u}r Experimentalphysik der Universit\"{a}t Wien, 1999
\end{thebibliography}
\end{document}